\documentstyle[12pt,aas2pp4]{article}
\begin{document}

\title{Inelastic clump collision model for non-Gaussian
velocity distribution in molecular clouds}
\author{Shigeru Ida}
\affil{Department of Earth and Planetary Science,
Tokyo Institute of Technology,
Oh-okayama, Meguro-ku, Tokyo 152, Japan}
\authoremail{ida@geo.titech.ac.jp}
\author{Y-h. Taguchi}
\affil{Department of Physics, Tokyo Institute of Technology,
Oh-okayama, Meguro-ku, Tokyo 152, Japan}
\authoremail{ytaguchi@cc.titech.ac.jp}
\begin{abstract}

Non-Gaussian velocity distribution in star forming region is
reproduced by inelastic clump collision model.
We numerically calculated the evolution of inelastic hard spheres
in sheared flow, which corresponds to cloud clumps in differential
galactic rotation.
This system fluctuates largely around equilibrium state,
creating clusters with inelastic collisions and destroying them
with shear motion.
The fluctuation makes spheres have non-Gaussian
velocity distribution with nearly exponential tail.
How far from Gaussian distribution depends upon
coefficient of restitution, which can produce the variety of degree of
deviation from Gaussian among regions.
\end{abstract}


\section{Introduction}

In general, the velocity distribution in molecular clouds has an excess high
velocity tail compared with Gaussian (e.g., \cite{Falgarone,Miesch}).
In particular, Miesch and Scalo
 pointed out that the most important
results are variety between subregions of molecular cloud
(in some subregions the distribution is
nearly Gaussian while it deviates significantly from Gaussian in
other subregions) and nearly exponential tail in the non-Gaussian cases
(\cite{Miesch}).

Several explanations for the non-Gaussian distribution have been proposed:
the high-velocity component came from unbound interclump medium
(\cite{Blitz}), high-speed clump collisions (\cite{Keto}),
nonlinear Alfven waves (\cite{Elmegreen}),
 or the intermittency of interstellar
turbulence (\cite{Falgarone}).
However, it is difficult for these mechanisms to explain the variety and
the exponential tail (Miesch and Scalo 1995).

We propose an alternative mechanism, clustering of clumps through
mutual inelastic collisions.
This mechanism causes the non-Gaussian distribution with the variety and the
nearly exponential tail with simple physics.
Molecular clouds have hierarchical structure.
Here we generically call density-contrasted areas with any sizes
^^ ^^ clumps''.
Clumps interact with one another through direct collisions and gravitational
scatterings.
In general, interparticle interactions in the external
gravitational potential converts
orbital energy to random motions (e.g., \cite{Goldreich,Ida,Ohtsuki}).
Sheared motion between particles,
which is due to differential rotation of galactic disk,
causes encounters between them, even if they have no random motion.
Thus random motions are produced by interclump collisions and/or gravitational
scatterings.
On the other hand, interclump collisions also have effect to dissipate
random motions, since they are supersonic, {\it i.e.}, inelastic.
Hence, some equilibrium state of random motions of clumps are maintained.

To make physics clear, we consider a simple model, inelastic hard spheres
in sheared motion and numerically calculated the dynamical evolution of
the system.
We neglected gravitational scattering, because what makes non-Gaussian
distribution is inelaticity of collisions (see below) and the effects of
gravitational scatterings are similar to those of elastic collisions (e.g.,
\cite{Cuzzi,Ohtsuki}).

As we will show, the fluctuation around the equilibrium state is large in the
case with relatively large inelaticity and densely packed spheres.
That is, the mean value of the velocity dispertion fluctuates considerably
and nonsteady spatial inhomogeneity of the distribution of the spheres
(^^ ^^ cluster'') is formed.
This behavior is characteristic of a system with both energy input and
dissipation and does not depend on details of dissipation mechanism.
Hence, the essential features of realistic clump collisions can be
described by our simple model, though dissipation is much more complicated
in realistic clump collisions.

The clustering of clumps produces non-Gaussian velocity distribution
with nearly exponential tail.
The deviation from Gaussian is closely related to the degree of clustering,
which is regulated by degree of inelasticity (restitution coefficients).
In this model, the variety is
naturally explained as different ^^ ^^ effective'' restitution coefficients
among subregions.

\section{Method of numerical calculation}

Numerical setup of our model is as follows.
We adopt rotating local Cartesian coordinates in a galactic disk.
Hard spheres with normal restitution coefficient $e$ are
packed into two dimensional box having width $L_y$ and height $L_x$.
The $x$ and $y$ directions correspond to radial and azimuthal directions of
the galactic disk, respectively.
We neglected ^^ ^^ softness'' of clumps, tangential restitution coefficient,
and the thickness
of the disk for simplicity, which hardly affects the properties of the
results here.

For the limitation of calculation, number of spheres, {\it i.e.,} $L_x$
and $L_y$, cannot be taken so large.
Hence the box corresponds to the area which is much smaller than the area
identified to one subregion by observation.
We describe dynamics of a whole subregion by time average of this box
instead of by summation of such boxes which cover a whole subregion.

This box has periodic boundary condition in the $x$ direction and
sheared boundary condition in the $y$ direction, corresponding to
sheared motion in a galactic potential (Fig.\ref{fig:sheared}).
When a sphere
passes through lower horizontal side of the box downward,
it re-enters into the box from upper horizontal boundary.
However, horizontal position differs between exiting position and
entering position.
Suppose that $y$ coordinate of the position where
the sphere goes out of box is $y_{out}$ and that of entering is $y_{in}$.
Then $y_{in}$ is equal to $y_{out}+ U t$,
 where and $t$ is time and $U$ is sheared velocity between
positions at upper and lower boundaries.
If $y_{out}+ U t$ is outside the box, $y_{in}$ is taken equal to be
$y_{out}+ U t - L_y$.
The $y$ component of velocity after re-entering box also increase
by $U; v_{in}=v_{out}+U$
where  $v_{in}$ and $v_{out}$
are the $y$ component of velocity when the sphere exits from
and re-enters the box respectively.
When a sphere passes through horizontal boundary upward,
$y_{out}$ should be $y_{in}-U t $
and $v_{in}=v_{out}-U$.
On the other hand, if a particle passes horizontally the vertical boundaries,
$x_{in} = x_{out}$ and $u_{in} = u_{out}$, where $u$ is the
$x$-component of velocity.

In addition to self-gravity of clumps, we neglected Coriolis force
and tidal force
(the difference between galactic potential force and centrifugal force)
in the equations of motion, because mean collision time is much shorter than
galactic rotation time, {\it i.e.}, characteristic time of Coriolis and
tidal forces.
(Preliminary results by 3D $N$-body calculation with self-gravity, Coriolis
and tidal forces show essentially the same results.)

\section{Results and their physical interpretation}

By using this model, we perform numerical simulations.
256 spheres with diameter 1.0 are considered.
The box size $L_x \times L_y$ is taken to be $18.2 \times 15.2$ and
sheared velocity $U=$ is 0.5.
The box size is taken to realize dense packing (in this case $\sim 70$\%).
Clusters of clumps are created by inelastic collisions and are smeared out
by sheared motion.
If packing is so sparse that mean collision time is longer than
smearing timescale, clusters are not maintained.
Hence dense packing is required to produce non-Gaussian distribution
due to clustering.
The choices of values of $L_x$, $L_y$, and $U$ do not change results,
as long as dense packing is realized.

Restitution coefficient $e$ is taken to be 0.6 and 0.9.
Result is sensitive to value of $e$.
After discarding transient period starting from initial condition with
equally spacing and random velocity ($\in [-0.05,0.05]$),
we measure  distribution of velocity component $u$\footnote{This
 is because distribution of $v$ component is strongly
deformed by shear velocity} over 120 (180) snapshots
for $e=0.9$ ($e=0.6$) every  256 (= total
number of spheres) collisions (Fig.\ref{fig:PDF_v})\footnote{Thus
 total number of collisions observed is
 $120\times256 =30720$ for
$e=0.9$ and 46080 for $e=0.6$.}.
The distribution in Figure\ref{fig:PDF_v}(a)
 clearly deviates from Gaussian distribution.
It has excess high-velocity tail that is nearly exponential.
This property is independent of initial conditions
when $e$ is relatively small.
This universal property is consistent with observations.
The degree of the deviation increases with deviation of the restitution
coefficient $e$ from unity.
The result with $e = 0.9$ shows almost Gaussian distribution
(See Fig. \ref{fig:PDF_v}(b)).

We found that non-Gaussian distribution is made by unsteady clustering
due to inelastic collision.
In Fig. \ref{fig:disp},
time evolutions of mean velocity dispersion are shown.
The vertical and horizontal axes are mean velocity dispersion and
cumulative collision numbers.
The velocity dispersion appreciably varies with time in the $e=0.6$ case.
In the $e = 0.9$ case, the variation is smaller.
In the limit of $e \rightarrow 1$, the velocity dispersion
fluctuates only within statistical one.
We found that this variance is induced by clustering of clumps.
Generally speaking, introduction of inelastic collision causes
spatial inhomogeneity. Goldhirsch and co-workers
(\cite{Goldhirsch,Goldhirsch2}) have
already pointed out that spatially homogeneous distribution of hard spheres
with inelastic collisions are unstable and collapse into several
clusters.
In our case, shear motion destroys the clusters.
Hence creation and destruction of clusters are repeated.
Through these processes, energy input from shear motion
is intermittently dissipated by inelastic collisions,
which makes large variation of velocity dispersion.

This time variance of dispersion does make
the deviation from Gaussian distribution, since
a composition of many Gaussian-like distributions with different dispersions
is non-Gaussian distribution with excess high-velocity tail.
Thus non-Gaussian velocity dispersion can be explained by nonsteady
clustering due to inelastic collisions.

As noted before, the time variance of the present calculation corresponds
to spatial variance in subregion of molecular cloud.
Hence, the observed non-Gaussian distribution with excess high-velocity tail
can be explained by spatial variance of clustering within a subregion
caused by inelastic interclump collisions.

This model explains the observed variety of the distribution
as the variety of ^^ ^^ effective'' restitution coefficient.
As shown in Fig. \ref{fig:PDF_v}, the deviation from Gaussian depends upon
restitution coefficient.
Effective restitution coefficient
would be different among subregions in clouds.
For example, effective $e$ can change with $\rho$ as follows.
The relative importance between inelastic collisions and gravitational
scatterings is indicated by the ratio of sizes $r$ of clumps
to their tidal (or Roche-lobe) radii $r_{t}$ defined by
$(GM/\Omega^2)^{1/3}$
where $M$ is clump mass and $\Omega$ is galactic rotational frequency;
if $r/r_{t} \ll 1$, gravitational scatterings are more important, and
vice versa (e.g., Ohtsuki 1993).
For clumps of molecular cloud in the neighborhood of the sun, we obtain
$$
\frac{r}{r_{t}} \simeq 0.1 \left( \frac{\rho}{10 M_{\odot} \mbox{pc}^{-2}}
\right)^{-1/3},
\eqno(1)
$$
where $\rho$ is mass density of clump.
This means that direct collisions are as effective as gravitational
scatterings in evolution
of random motion of clumps, analogous to planetary rings.
Since gravitational scattering is equivalent to $e =1$ collisions,
effective $e$ can change with $\rho$ through Eq.(1).
Furthermore, different configurations and random velocities of clumps may
change effective $e$.

Thus inelastic interclump collisions explain the nearly exponential
high-velocity tail and the variety of distribution, and hence it
can be a basic mechanism
why velocity distribution in molecular cloud deviates from Gaussian.

\acknowledgments

We would like to thank Hideki Takayasu for fruitful discussion.

\clearpage
\figcaption{Local Cartesian coordinates with sheared boundary condition.
\label{fig:sheared}}
\figcaption{Velocity distribution (a) $e=0.6$ (b) $e=0.9 $
$u$ is renormalized so as to have variance of unity.
\label{fig:PDF_v}}
\figcaption{Time evolutions of mean velocity $u$ dispersion are shown.
Upper: $e=0.9$, Lower: $e=0.6$.
Data before 5120th collison are discarded.
\label{fig:disp}}
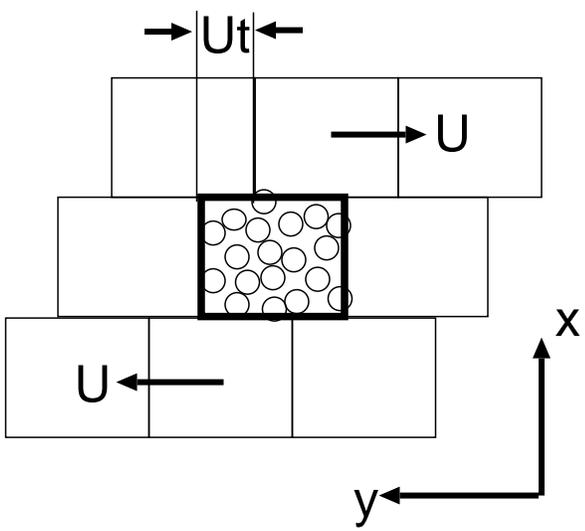
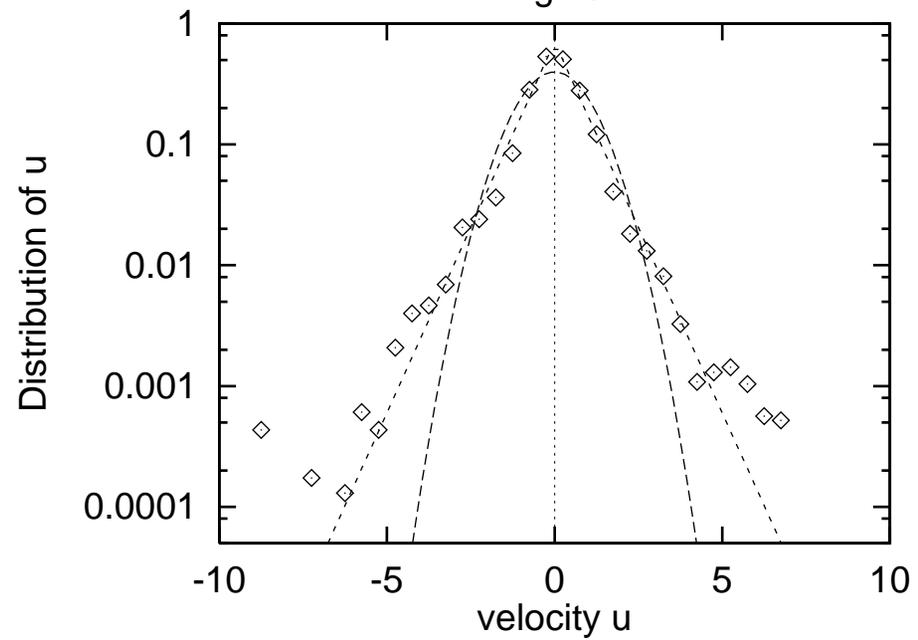
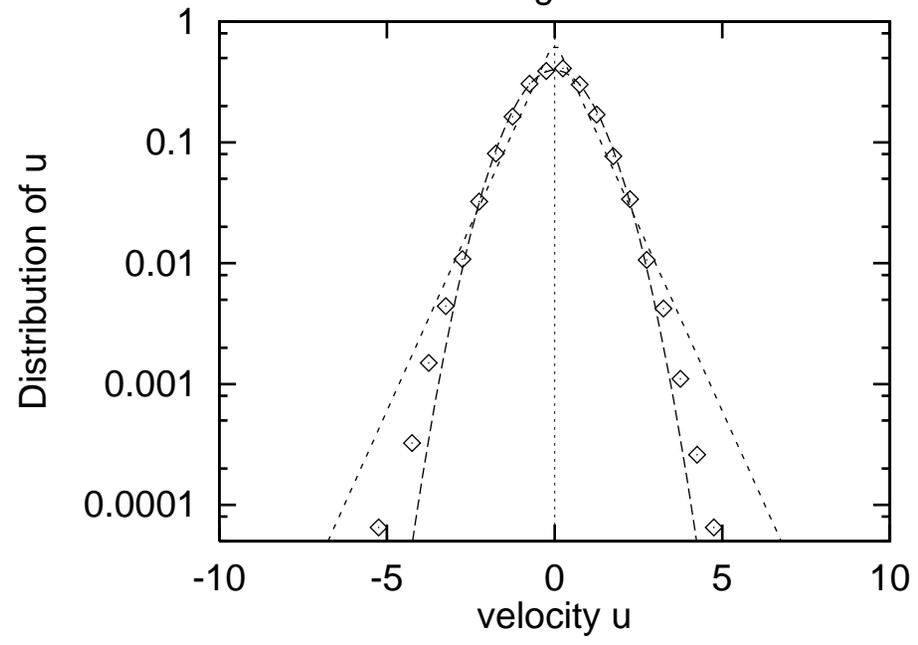
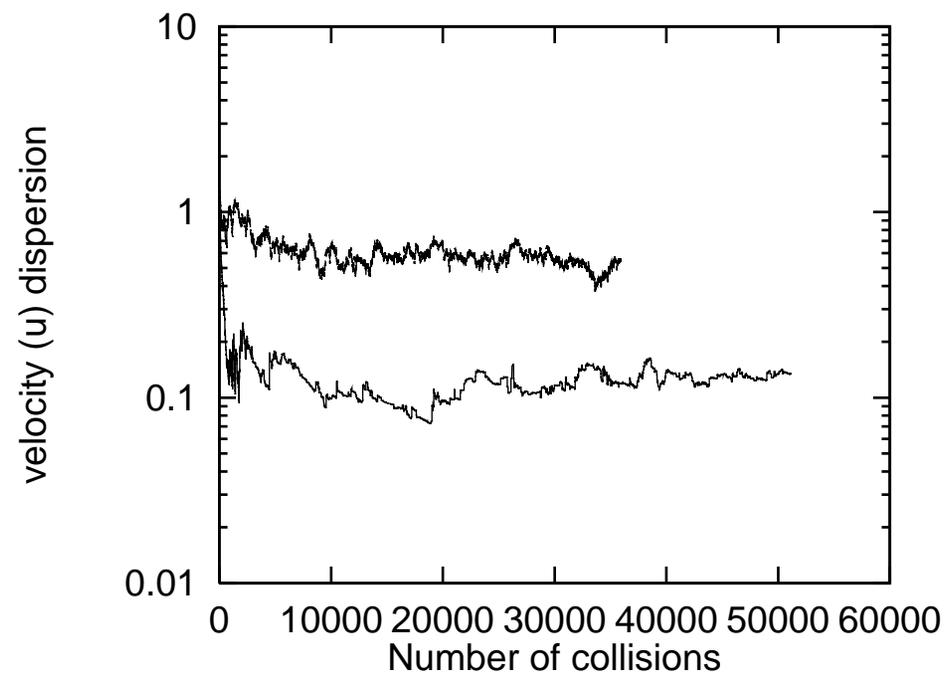
\end{document}